# Pulling Radiation Force State on Active Spherical Carriers via Varying Phase Incident Wave Field

Majid Rajabi[a,1], Hossein Khodavirdi[b,2]

[a] *School of Mechanical Engineering, Iran University of Science and Technology, Tehran, Iran*

[b] *School of Mechanical, Manufacturing and Aerospace Engineering, Illinois Institute of Technology, Chicago, Illinois, USA, 60616*

On the way of investigating a complete maneuverability technique and based on the emergence of negative acoustic radiation force on an active carrier in a progressive plane wave field, we propose a new technique in which not only the attractive force is experienced but also the average of acoustic radiation force over the location of the object with respect to the sound source is negative. Keeping in mind that the produced pulling force upon a carrier activated on only one mode always have a positive average, along with the point that the positive average force means that the net motion will be pushing in lieu of a pulling effect associated with a negative force, made us to put forward a technique to generate the acoustic radiation force with negative average over a complete wavelength to ensure that the net motion is attractive.

The idea here is to excite the object in two modes (zero and first) synchronously with a varying difference between the phases of excitations on two modes with the hope of generating acoustic radiation force with a negative average over a complete wavelength.

**Keywords:** *Maneuverability; Negative Average Force; Acoustic Radiation Force; Pulling effect.*

---

[1] Sustainable Manufacturing systems research laboratory, Email of Corresponding Author: majid_rajabi@iust.ac.ir
[2] Wave Research Laboratory, Email: hkhodavirdi@hawk.iit.edu

# I. Introduction:

The problem of manipulation of small objects via acoustic radiation force and torque caused by a single frequency acoustic wave on objects with different shapes has inspired a large body of scientific research. The paper of King (Vessot, 1934) on 1934 which can be considered as the cornerstone of this concept, proposed a formulation for the acoustic radiation pressure on rigid spheres in non-viscous fluids. Hasegawa and Yosioka also propounded their formulation for acoustic radiation pressure on elastic solid spheres in a plane progressive wave (Hasegawa and Yosioka, 1969; Hasegawa, 1977). Acoustic radiation force also was of concern in works like (Westervelt, 1951; Hasegawa and Yosioka, 1969; Torr, 1984) which they found the acoustic radiation force with slightly different approaches for objects of different shapes. Moreover, major contributions were done on acoustic radiation force and torque in progressive (Silva, 2011; G. T. Silva, 2013; Silva, 2014; G. T. Silva, 2018) and standing wave fields (Kozuka *et al.*, 2008; Liu and Hu, 2009; G. T. Silva, 2018) for different object shapes (Hasegawa *et al.*, 1988; Lima *et al.*, 2020). The application of acoustic field was also evaluated in areas like levitation (Foresti *et al.*, 2011; Lim *et al.*, 2019a; Lim *et al.*, 2019b), sorting (A. Haake, 2004), focusing (Liu and Hu, 2009; Foresti *et al.*, 2011; Courtney *et al.*, 2014) etc.

The recent efforts on developing the concept and idea of active carriers on the path of promoting the acoustic handling technique, led to the introduction of the new concepts of acoustically activated bodies as self-motile carriers (Mojahed and Rajabi, 2018; Rajabi and Hajiahmadi, 2018), self-activated carriers manipulated with the monochromatic incident beam (Rajabi and Mojahed, 2016a; b; 2017a; b; 2018), or externally activated bodies with compound fields (Yu *et al.*, 2018a; Yu *et al.*, 2019) and just recently, it is shown that how 3D steering may be achieved via a bi-polar configuration of the internal excitation of a spherical active carrier (Rajabi *et al.*, 2020). It is shown that the distortion of the wave field may lead to the generation of interesting negative radiation force or pulling effects (Marston, 2007; Rajabi and Mojahed, 2016b; Yu *et al.*, 2018a) in association with the positive radiation force or pushing effects. To be more specific, using the complex phase shift and by highlighting the relation of power absorption with acoustic radiation force, Marston and Zhang (Marston and Zhang, 2016; Zhang and Marston, 2016) showed that for a passive object in a plane progressive wave field it is not possible to have negative absorbed energy or negative acoustic radiation force. Nevertheless, due to the point that scattering in the backward hemisphere is negligible in the Bessel beam case comparing to the simple plane wave case, the same passive object in a focused acoustic field like Bessel beam experiences negative acoustic radiation force at specific frequencies and cone angles (Marston, 2006).

Making a review on the generated negative radiation force on the active carriers (Marston, 2007; Mitri, 2009; Yu *et al.*, 2018b), it is found that for the single canonical mode of excitation in the case of plane progressive wave field, as the most popular and practical wave form, both states of positive and negative radiation forces exist. So, similar to the case of standing wave field on the passive carriers, in the range of a wavelength, here also two stable and unstable rest states (i.e., zero radiation force) induced so that the carrier will pull toward the stable one (Rajabi *et al.*, 2020). The suggested manipulation strategy is the change of phase difference between the incident wave and the radiated field (the object), so that the location of the stable rest point has been varied and the carrier will be manipulated. It must be noticed that for a single-mode excited carrier the average of the acoustic radiation force over a wavelength is same as the acoustic radiation force on the carrier on its passive (no-radiation) state.

In the present work, another idea of manipulation strategy is pursued focusing on the concept of generating net negative radiation force independent of the location of the carrier. The active carrier is stimulated on its two first modes of vibration (i.e., monopole and dipole modes) with specified phase difference and the same frequency. The formulation section of this paper starts with a brief introduction to the configuration of the problem and it continuous with the acoustic field equations in sub-section B along with the calculations of scattered coefficients in sub-section C. Section II ends by two final sub-sections, first a mathematical manipulation on the acoustic radiation force formulation in sub-section D and then a quick review on the dynamics of a moving particle in a fluid in sub-section E. In the results and discussion section, by giving a numerical example, first it is shown that the MATLAB code which is written based on the formulations are valid for the case of a single-mode activated sphere, then Fig. 3 shows that for objects which are activated in a single mode, the average of the force over the location of the object is always positive and equal to the acoustic radiation force (ARF) acting on a passive object and then it is shown that for a wide range of phase differences between two excitation modes of the carrier, at some specified frequencies, the net negative radiation force is achievable. In the continuation of the result section, by considering the dynamics of the carrier in the host fluid medium and assuming the linear time variation of the phase of incident wave field, it is shown that four states of the radiation force are possible:

- The positive average of radiation force with always positive values of RF.
- The positive average of radiation force with positive/zero/negative values of RF on a wavelength and induction of stable/unstable rest states.
- The negative average of radiation force with always negative values of RF.
- The negative average of radiation force with positive/zero/negative values of RF on a wavelength and induction of stable/unstable rest states.

Taking into account the dynamics of the active carrier in the host medium and by keeping in mind the existence of the drag force, the history force and the added mass effects (Happel and Brenner, 1983), the time dependency of traveling velocity of the carrier is calculated and it is observed that for the rate of phase change greater than a (frequency dependent) minimum value, the traveling velocity of carrier approaches the stable value, backward the incident wave field direction.

Although the empirical aspect of the design and fabrication of active carriers using piezo ceramics has not been seen by the authors yet, there are many published papers based on practical experiments on acoustic manipulations of passive particles using different acoustic fields (S. A. Seah, 2014; Lim *et al.*, 2019a; Lim *et al.*, 2019b). Moreover, Hai-Qun Yu proposed a method (Yu *et al.*, 2019) which seems feasible to be tested practically which it is about exciting liquid spheres by a laser beam instead of using piezo ceramics to generate negative acoustic radiation force, albeit that work was again a theoretical study. The authors hope that the present work in addition to other recently introduced ideas may make a tunnel through the novel concept of active carriers and acoustic manipulation.

## II. Formulation

### A. Configuration

Here, an exact definition of the problem is introduced. The general shape of the carrier is presented in Fig. 1. In the 2D plane of view, the horizontal axis is considered to be *Z* and the vertical one to be *X* and as it is shown the spherical coordinates are *r* and $\theta$.

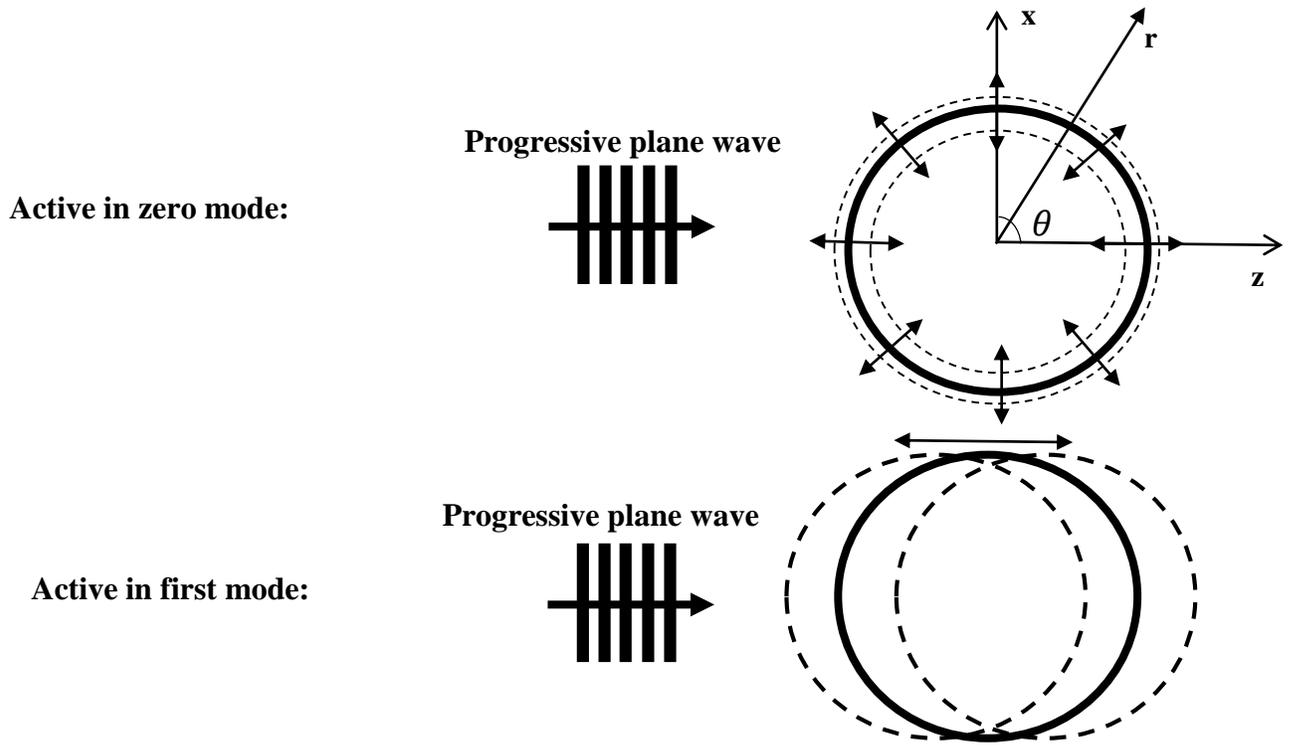

**FIG. 1.** This figure shows a schematic of the geometry and the excitation of the sphere in its zero and first mode.

Moreover, the incident plane wave propagation's direction is on *Z* axis with no angular deviation. The spherical object has a two-layer structure that its outer layer is an elastic casing with inner and outer radius of *b* and *a* respectively. The inner layer is a piezoelectric ceramic with the inner and outer radius of *c* and *b* respectively and it is to excite the object on its monopole (breathing) and dipole modes of vibration by using a same-amplitude and different phases.

## B. Acoustic field equations

Considering the time-independent form of the wave equation (Skudrzyk, 2012), we have:

$$(\nabla^2 + k_f^2)\varphi = 0 \tag{1}$$

where $k_f = \omega/c_f$ is the wave number and $\varphi$ is the velocity potential of the acoustic wave in external fluid. Based on the axisymmetric nature of the problem and knowing that the velocity potential is a function of four parameters ($r, \theta, \omega$ and $t$), the solution for the Helmholtz partial differential equation would be:

$$\varphi_{inc}(r,\theta,\omega) = \varphi_0 \sum_{n=0}^{\infty}(2n+1)i^n j_n(k_f r) P_n(\cos\theta) e^{-i\omega t} \tag{2}$$

where $j_n$ is the spherical Bessel function of the first kind of order n, $P_n(\cos\theta)$ are Legendre polynomials, and $e^{-i\omega t}$ is the harmonic time variation. Keeping in mind the relation of pressure field with the velocity potential in fluid, gives:

$$p(r,\theta,\omega) = \rho_f \frac{\partial \varphi}{\partial t} \tag{3}$$

So, we will have the pressure field of incident acoustic wave like:

$$p_{inc.}(r,\theta,\omega) = -\varphi_0 i\omega\rho_f \sum_{n=0}^{\infty}(2n+1)i^n j_n(k_f r) P_n(\cos\theta) e^{-i\omega t} \tag{4}$$

Likewise, the solution for the Schrodinger equation for the scattered velocity potential function in the surrounding medium can be written as a combination of spherical waves. It is also essential to keep in mind the Sommerfeld radiation condition for the scattered field.

$$\varphi_{Scatt.}(r,\theta,\omega) = \varphi_0 \sum_{n=0}^{\infty} A_n(\omega)(2n+1)i^n h_n^{(1)}(k_f r) P_n(\cos\theta) e^{-i\omega t} \tag{5}$$

Where $h_n^{(1)} = j_n + iy_n$ is the spherical Hankel function of the first kind of order n and $A_n$ are the unknown modal scattering coefficients which for the case that the object is passive is determined by employing the appropriate boundary conditions on (Rajabi and Mojahed, 2016a). Based on the superposition in linear acoustic regimes, the total velocity potential and the total acoustic pressure in surrounding medium can be written like:

$$\varphi_{total} = \varphi_{Scatt.} + \varphi_{inc.}, \tag{6}$$

$$p_{total}(r,\theta,\omega) = \rho_f \frac{\partial \varphi_{total}}{\partial t}. \tag{7}$$

**C. Scattered coefficients and imposed voltage:**

Based on what is indicated on (Rajabi and Mojahed, 2016a) for finding the scattered coefficient by using the appropriate boundary conditions like the equality of fluid pressure and the radial stress of the object at $r = a$ or equality of the medium displacement and the elastic layer radial displacement at $r = a$ for the case of a two-layer spherical object we can have $A_n(\omega)$ for an active sphere and concurrently by using the Cramer rule $A_n(\omega)$ can be rewritten in a way that the passive and active terms be distinctly present. So, we have:

$$A_n(\omega) = Z_1^n + Z_2^n \Phi_n, \tag{8}$$

where $Z_1^n$ and $Z_2^n$ are the modal transfer functions of the system present in the appendix and for a detailed version readers are referred to (Rajabi and Mojahed, 2016a). Also, $\Phi_n$ is the modal voltage amplitude applied to the piezoelectric actuator which is obtained from the modal expansion of the imposed electric potential amplitude and the boundary condition which is about prescribed axisymmetric electrical voltage imposed at the inner and outer surfaces of the piezo actuator.

$$\Phi(r = b, \theta, t) = V(\theta, \omega)e^{-i\omega t}, \tag{9}$$

$$\Phi(r = c, \theta, t) = 0. \tag{10}$$

Rewriting the modal expansion for the imposed electric potential amplitude brings about two expressions for the modal voltage amplitude for the zero and first modes.

$$V(\theta,\omega) = \frac{e_{33}a}{\varepsilon_{33}} \sum_{n=0}^{\infty} \Phi_n P_n(\cos\theta), \qquad (11)$$

Taking advantage of the orthogonality property of the Legendre polynomials which is:

$$\int_{-1}^{+1} P_n(x) P_m(x) dx = \frac{2}{2n+1} \delta_{nm}, \qquad (12)$$

where $\delta_{nm}$ is the Kronecker delta and it is equal to 1 when $n = m$ and otherwise it is equal to 0. The mentioned property gives us the opportunity to multiply both sides of Eq. (11) by $-P_n(\cos\theta)\sin\theta d\theta$, Then we have:

$$\frac{\varepsilon_{33}}{e_{33}a} \int_0^{\pi} V P_n(\cos\theta)\sin\theta d\theta = \int_0^{\pi} \Phi_n P_n(\cos\theta) P_n(\cos\theta)\sin\theta d\theta, \qquad (13)$$

Based on the raised challenge that whether we can have negative values for the average of force over the phase difference of the incident wave and monopole excitation or not, and by considering that the average of force on the mentioned phase difference is always positive when only the zero mode is active, we suggest to activate the monopole and dipole modes simultaneously with the same value for voltage amplitude, but a deviation in phase. Therefore, the imposed voltage on the object is $V = V_0 + V_1 \cos\theta$. Now by using Eq. (12) and keeping in mind that the voltage in the dipole mode is dependent on $\theta$ but it is independent for the breathing mode, the expression for the modal electric potential amplitude will be:

$$\Phi_n = \left(\frac{2n+1}{2}\right) \frac{\varepsilon_{33}}{e_{33}a} \int_{-1}^{1} V P_n(x) dx, \qquad (14)$$

For the modal electric voltage for the monopole and dipole modes, we found $\Phi_0 = (\varepsilon_{33}/e_{33}a)V_0$ and $\Phi_1 = (\varepsilon_{33}/e_{33}a)V_1$ where $V_0 = V_{amp}e^{-i\phi_0}$ and $V_1 = V_{amp}e^{-i(\phi_0+\Delta\phi)}$ in which $V_{amp}$ is the amplitude of imposed voltage for both modes, $\Delta\phi$ indicates the phase difference between the excited modes and $\phi_0$ is the phase difference between the monopole mode of excitation and the incident wave filed.

Now, the expressions for the scattered coefficients are found as

$$\begin{cases} n=0, \to A_0 = Z_1^0 + Z_2^0 \dfrac{\varepsilon_{33}}{e_{33}a} V_{amp} e^{-i\phi_0} \\ n=1, \to A_1 = Z_1^1 + Z_2^1 \dfrac{\varepsilon_{33}}{e_{33}a} V_{amp} e^{-i(\phi_0+\Delta\phi)} \\ n \geq 2, \to A_n = Z_1^n \end{cases} \qquad (15)$$

**D. Acoustic radiation force based on the strategy:**

The time-averaged radiation force function is represented like:

$$\langle F \rangle_t = E_{inc} S_c Y_z, \qquad (16)$$

where $E_{inc} = \rho_f k_f^2 \varphi_0^2 / 2$ is the incident wave energy density and $S_c = \pi a^2$ is cross-sectional area of the spherical object. Moreover, $Y_z$ is the non-dimensional form of acoustic radiation force in Z- direction as (Rajabi and Mojahed, 2016a; Zhang and Marston, 2016):

$$Y_z = \left[ -4/(k_f a)^2 \right] \sum_{n=0}^{\infty} (n+1)\left[ \alpha_n + \alpha_{n+1} + 2(\alpha_{n+1}\alpha_n + \beta_n \beta_{n+1}) \right], \qquad (17)$$

where $\alpha_i$ and $\beta_i$ are real and imaginary parts of scattering coefficients, respectively. Here, by using the Eq. (8) and after some manipulations the expression for $\alpha_i$ and $\beta_i$ for $i = 0$ and 1 can be written like

$$\alpha_0 = \text{Re}(Z_1^0) + J\left[V_{amp}\text{Re}(Z_2^0)\cos\phi_0 + V_{amp}Im(Z_2^0)\sin\phi_0\right]$$

$$\beta_0 = Im(Z_1^0) + J\left[-V_{amp}\text{Re}(Z_2^0)\sin\phi_0 + V_{amp}Im(Z_2^0)\cos\phi_0\right]$$

$$\alpha_1 = \text{Re}(Z_1^0) + JV_{amp}\cos\phi_0\left[\text{Re}(Z_2^1)\cos\Delta\phi + Im(Z_2^1)\sin\Delta\phi\right]$$
$$+ JV_{amp}\sin\phi_0\left[Im(Z_2^1)\cos\Delta\phi - \text{Re}(Z_2^1)\sin\Delta\phi\right] \quad (18)$$

$$\beta_1 = \text{Re}(Z_1^1) + JV_{amp}\cos\phi_0\left[Im(Z_2^1)\cos\Delta\phi - \text{Re}(Z_2^1)\sin\Delta\phi\right]$$
$$+ JV_{amp}\sin\phi_0\left[-\text{Re}(Z_2^1)\cos\Delta\phi - Im(Z_2^1)\sin\Delta\phi\right],$$

where, Re(.) and Im(.) are the real and imaginary parts of (.), respectively and $J = \varepsilon_{33}/e_{33}a$.

Regarding the active and passive modes, the expression for the non-dimensional force can be re-written like:

$$Y_z = \left[-4/(k_f a)^2\right]\left\{\sum_{n=2}^{\infty}(n+1)\left[\alpha_n + \alpha_{n+1} + 2(\alpha_{n+1}\alpha_n + \beta_n\beta_{n+1})\right]\right.$$
$$\left. + \left[\alpha_0 + 2\alpha_0\alpha_1 + 2\beta_0\beta_1 + \alpha_1(3+4\alpha_2) + 2\alpha_2 + 4\beta_1\beta_2\right]\right\},$$

(19)

After some manipulations, the expression for $Y_z$ can be re-written in the form:

$$Y_z = \left[-4/(k_f a)^2\right]\left(D\cos^2\phi_0 + E\sin^2\phi_0 + F\sin\phi_0\cos\phi_0 + G\cos\phi_0 + H\sin\phi_0\right) + Y_z^{Passive}, \quad (20)$$

where $Y_z^{Passive}$ is the passive radiation force, $D$, $E$, $F$, $G$ and $H$ are expressions which are functions of $\Delta\phi$. Knowing that the integral of $\sin\phi_0\cos\phi_0$, $\cos\phi_0$ and $\sin\phi_0$ over a complete wavelength or over $[0, 2\pi]$ is zero, leads to the following expression for the averaged non-dimensional force over $\phi_0$ as,

$$\langle Y_z \rangle_{\phi_0} = Y_z^{Passive} + \left[-8/(k_f a)^2\right](JV_{amp})^2 \left\{\cos\Delta\phi\left[\text{Re}(Z_2^0)\text{Re}(Z_2^1) + \text{Im}(Z_2^1)\text{Im}(Z_2^0)\right]\right.$$
$$\left. + \sin\Delta\phi\left[\text{Re}(Z_2^0)\text{Im}(Z_2^1) - \text{Re}(Z_2^1)\text{Im}(Z_2^0)\right]\right\}, \quad (21)$$

where,

$$Y_z^{Passive} = \left[-4/(k_f a)^2\right]\left\{\sum_{n=2}^{\infty}(n+1)\left[\alpha_n + \alpha_{n+1} + 2(\alpha_{n+1}\alpha_n + \beta_n\beta_{n+1})\right] + \right.$$
$$\left. + \text{Re}(Z_1^0) + 2\text{Re}(Z_1^0)\text{Re}(Z_1^1) + 2\text{Im}(Z_1^0)\text{Im}(Z_1^1) + (3+4\alpha_2)\text{Re}(Z_1^1) + 2\alpha_2 + 4\beta_2\text{Im}(Z_1^1)\right\}. \quad (22)$$

### E. Dynamics of Carrier

In most of the original works done in acoustic manipulation area of research, the attentions were mostly on steady solutions of the problems to improve and complete the concepts along with adding mathematical methods to enrich the topic of acoustic manipulation. Nevertheless, to cultivate and support the concept, the correlation between the rate of change in particle phase and the velocity of carrier due to the effect of acoustic radiation force, Stokes drag, added mass force and Basset force is investigated. The equation of motion of a particle in viscous fluid is like(J. Happel, 1983),

$$m_c \frac{dV_c}{dt} = -6\pi\eta a V_c - \frac{1}{2}m_f \frac{dV_c}{dt} - 6\pi\eta a^2 \int_0^t d\tau \frac{dV_c/d\tau}{[\pi\nu(t-\tau)]^{1/2}} + F_{acoustic}, \quad (23)$$

The first, second and third terms in right hand side of Eq. (23) are Stokes drag force, added mass force and history force respectively and the last term here is acoustic radiation force. For this equation, $V_c$ is the velocity of carrier, $\eta$ is the dynamic viscosity, $\nu$ is the kinematic viscosity, $m_c$ is the mass of carrier and $m_f$ is the mass of the displaced fluid. Here, we used Laplace transform to solve Eq. (23) and find the expression for velocity of particle. Due to the point that Basset force has a meaningful magnitude just when the body accelerates with high rate and considering the point that its existence in the equation makes the mathematical manipulation cumbersome in vain, along with the point that we are just looking for an approximate behavior of the body, we neglect the effect of history force(Johnson, 1998). After implementing Laplace transform, we will have:

$$\left[(m_p + m_f/2)s + 6\pi\eta a\right]\bar{V}_c = \bar{F}_{acoustic}, \tag{24}$$

Here, $\bar{(.)}$ means the transformed function. The final expression for the velocity of particle will be $v_i = f * g = \int_0^t F_i(t-\tau)g(\tau)d\tau$, where $g$ is $g = L^{-1}\left\{1/\left[(m_p + m_f/2)s + 6\pi\eta a\right]\right\}$, $F$ is the acoustic radiation force and $L^{-1}\{.\}$ means Laplace inverse of the function.

## III. Results and Discussion:

To assess the validity of the derived expressions, assumptions and formulae, a numerical example was considered and solved via a written code in MATLAB. The surrounding fluid is assumed to be water with the pressure equal to 100 kPa along with the sound speed of 1497 m/s. On the assumed condition, the fluid density is 997 kg/m$^3$. Moreover, the 1-mm-radius object was modeled to be a two-layer sphere with a layer of piezoelectric with the thickness of 0.1 mm which is fabricated from PZT4 and a 0.1-mm isotropic stainless-steel layer which performs as a coating. The properties of materials are mentioned in (Rajabi and Mojahed, 2016a). In previous works, although the efforts to excite the object on its breathing mode in one work (Rajabi and Mojahed, 2016a) and on its first mode in another work (Rajabi and Mojahed, 2018) led to valuable results about cancellation voltage which helps us to determine a logical and practical value for the operational voltage for this work, the outcome about generation of negative force average over the position of object (or the phase difference with transducer) was dissatisfying. Here, the operational voltage was decided to be 2 times of cancellation voltage defined as the minimum required voltage amplitude in order to achieve the zero-radiation force state for the case of interaction of an activated spherical object in its breathing mode and a plane progressive acoustic wave (Rajabi and Mojahed, 2016a). For assurance, the results were examined for other multipliers of cancellation voltage. Fig. 2a and Fig. 2b show the minimum normalized voltage amplitude and phase for cancelling the acoustic radiation force vs. $ka$ (the non-dimensional frequency), respectively, for the cases which breathing, and first modes were individually excited. The conformity of these figures and figures 4(a) and 4(b) in (Rajabi and Mojahed, 2016a) can verify our future research on activation of the modes individually (not simultaneously) and also, they can be considered as a validation of the operational voltage.

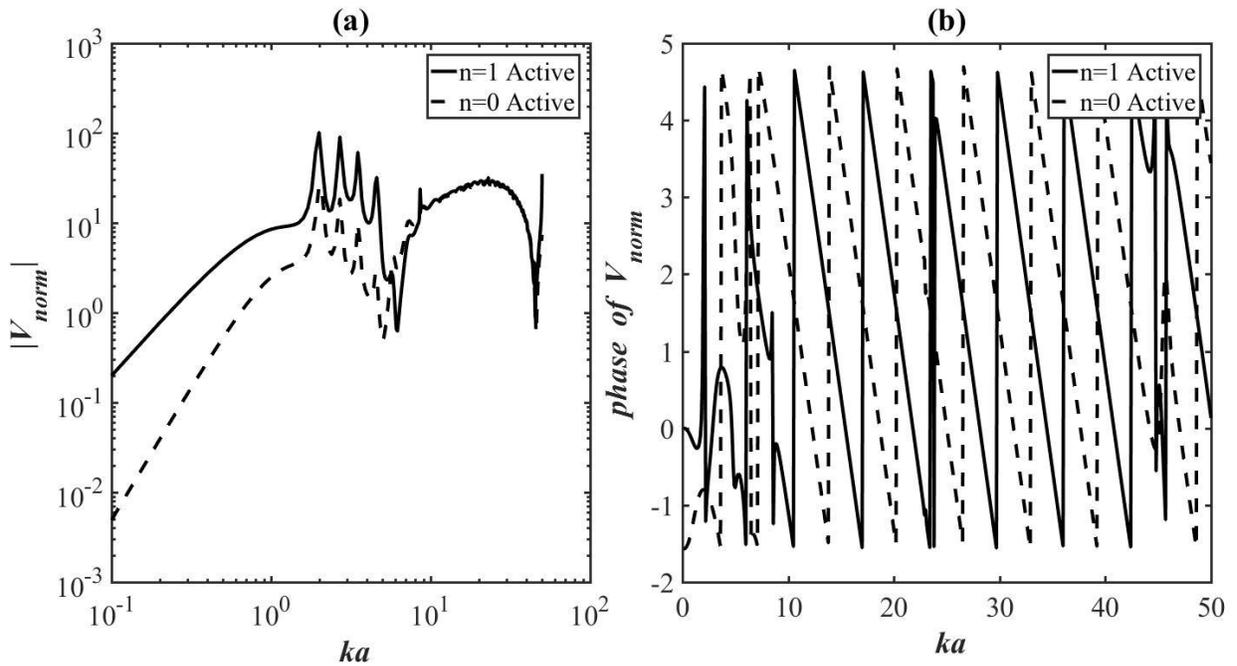

**FIG. 2.** (a) The amplitude of minimum normalized cancellation voltage and (b) the phase of minimum normalized cancellation voltage Vs. non-dimensional frequency. ($n=0$ or $n=1$ is active for the carrier)

As it can be proven mathematically that the average of force over the relative position of object in one complete wavelength is always positive for the case which the sphere is active in its first mode and taking into consideration this finding proved by a numerical example in (Rajabi and Mojahed, 2018), in this work, the mentioned theory is also tested mathematically and also by using a same numerical example for the case of an active carrier in breathing mode. Fig. 3 shows non-dimensional force in Z-direction for the interaction of a spherical carrier for two cases of excitation in breathing and monopole modes, individually. Here, the operational voltage is again 2 times of the cancellation voltage for a carrier vibrating in breathing mode and the operational non-dimensional frequency is $ka=2$ which is approximately 470 kHz for a carrier with outer radius of 1 mm. An identical positive value of 0.9252 for the average of the non-dimensional force function over the phase is manifested for both cases, equal to the radiation force exerted on the carrier in the case of a passive state.

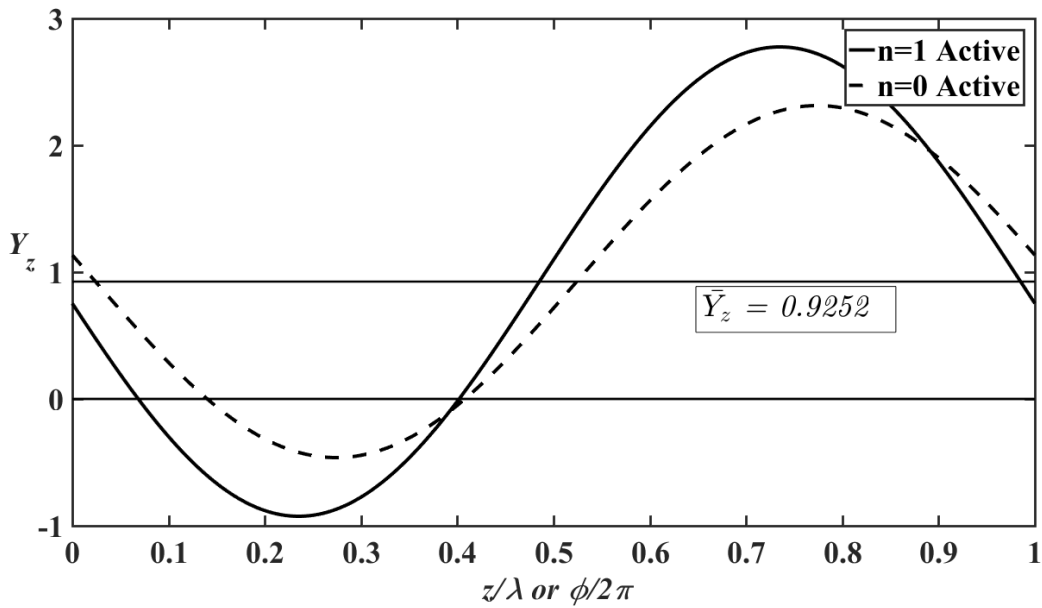

**FIG. 3.** The non-dimensional force in Z-direction versus phase difference between the incident wave field and the activated carrier at $ka=2$. ($n=1$ or $n=0$ is active for the object.)

In order to show the practicality of the derived formulas, it is time to discuss about figures which illustrate the average of force over the position of the object relative to the source (or its phase relative to the phase of incident wave) as a function of two efficacious parameters, frequency and the modal phase difference, defined as the phase difference between the vibrations of the shell in breathing and dipole modes. These parameters called efficacious as they are well-adjustable and if the results show wide intervals for the frequency and the phase-difference which the average of force is negative on those conditions, a practical system can be designed to work with a determined frequency and phase difference which brings about negative values for the average of force. Confidently, the desired occasion is happened for a vast number of practical frequencies and phase-differences which can be concluded from Figs. 4a and 4b. After plotting these figures for a practical range of frequency, $0.1 \leq ka \leq 50$ (approximately equal to $24\ kHz \leq f \leq 12\ MHz$), it was concluded to demonstrate it in the range of $2 \leq ka \leq 5.3$ (approximately equal to $470\ kHz \leq f \leq 1.2\ MHz$) for two purposes:

- The first frequency which the average of non-dimensional acoustic radiation force touches negative values for wide intervals of phase-difference is $ka \approx 2$.
- The values for the average of the non-dimensional acoustic radiation force for frequencies over 1.2 MHz are getting big numbers which it makes the zones of negative values indistinct for lower ranges of frequency in the 3D plot and in the contours.

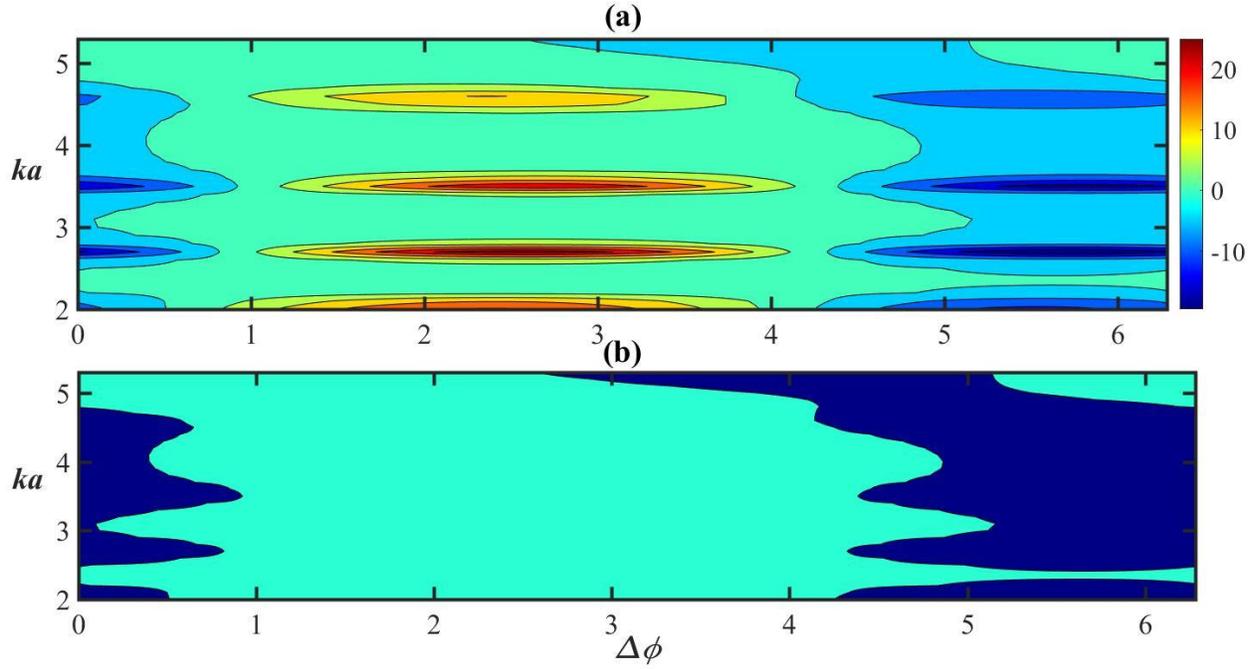

**FIG. 4.** (a) The 2D colored contour plot of the averaged non-dimensional force over the phase of incident wave (i.e., over the wavelength of the incident wave field), versus the non-dimensional frequency and the modal phase difference (i.e., the phase difference between the excitation of monopole and dipole modes of vibration) and (b) the 2D contour to show the islands of negative and positive values for the averaged non-dimensional acoustic radiation force. Islands with dark color are representing the negative values and the those with light color are for the positive ones.

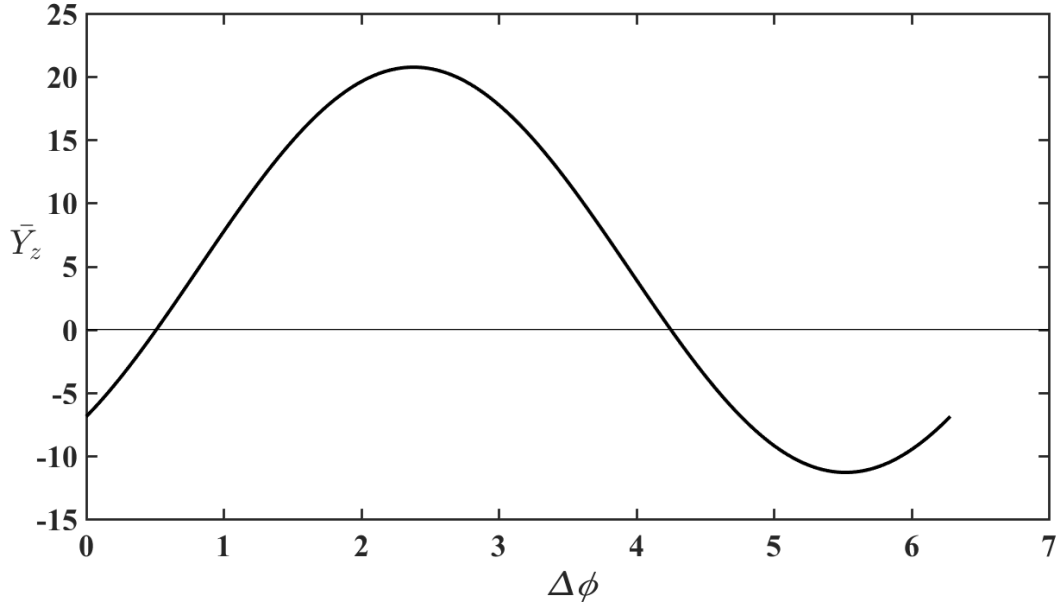

**FIG. 5.** A slice of the 3D plot of averaged non-dimensional acoustic radiation force versus modal phase difference, $\Delta\phi$, at selected non-dimensional frequency, $ka = 2$.

Fig. 4a depicts the 2D colored plot of the averaged non-dimensional acoustic radiation force in the plane of $ka - \Delta\phi$. Fig. 4b is another representation of Fig. 4a, separating the zones of negative and positive averaged radiation force states which the zones with dark color indicate the negative (pulling) state and the zones with light color indicate the positive (pushing) state. The black lines between the positive and negative force states presents the zero-averaged force states (i.e., $\bar{Y}_z = 0$).

Fig. 5 demonstrates a slice of Fig. 4, at selected frequency of $ka = 2$. As it is seen in the illustrated figure, many zones may be found that the interesting negative averaged radiation force is being generated, making possible the pulling states of radiation force. Furthermore, the special case of zero averaged radiation force state is occurred for two values of modal phase difference.

Keeping in mind the always positive averaged value of the acoustic radiation force over a wavelength of the incident wave, $z/\lambda$, or over the phase difference between the incident wave field and the radiated field of the carrier in its single mode of vibration (i.e., monopole or dipole mode), $\phi_0/(2\pi)$, this time for the case which the object is excited in two modes, Fig. 6 illustrates the acoustic radiation force, $Y_z$, as a function of $z/\lambda$ or $\phi_0/(2\pi)$, for selected values of modal phase difference, $\Delta\phi$, at the selected frequency of $ka = 2$. The sinusoidal pattern of the radiation force function over the phase difference or spatial position, generated in different states like big positive averaged forces, big negative averaged forces and small positive or negative averaged forces are present in Fig. 6. According to Fig. 4. it is understood that choice of the modal phase difference at a specific frequency determines whether the averaged force is a positive number or negative and also based on the magnitude of that averaged force one can guess whether the radiation force only takes positive values (means always pushing state), only negative values (means always pulling state) or like the previously reported case, it takes both positive and negative values. The zero force state occurs in the last case and two rest states (i.e., $Y_z = 0$) happens, one of them as the stable rest state and the other as the unstable rest state, which both are indicated in the figure as *S* and *U*, respectively (Rajabi and Mojahed, 2018).

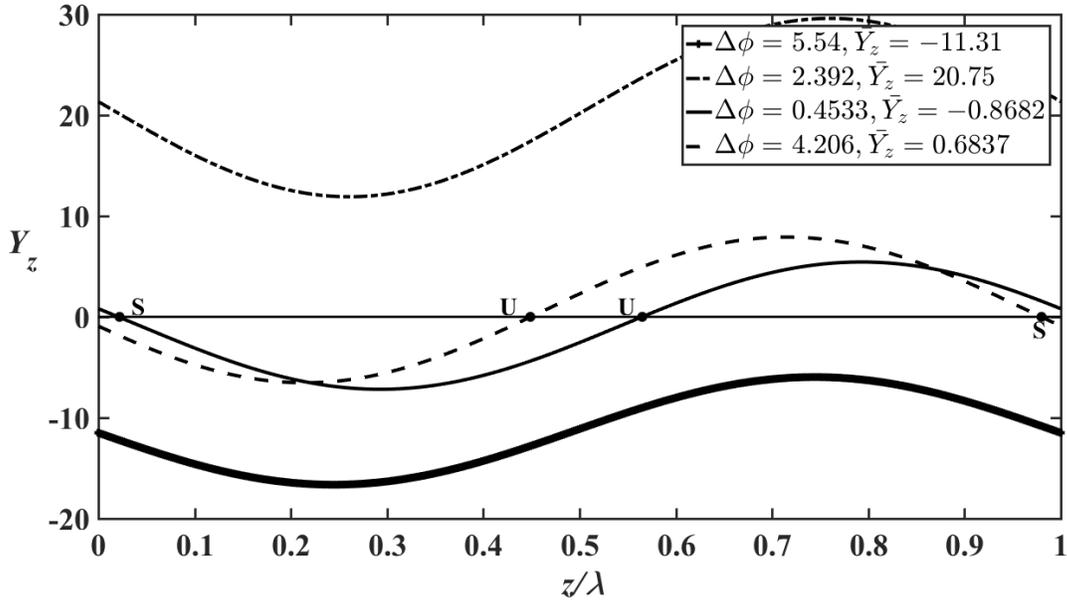

**FIG. 6.** This figure shows the non-dimensional force in z direction versus the phase difference of carrier and the wave (location of carrier) in four different states: I. Average of force is negative and the force is always negative, II. Average of force is negative but the force has both negative and positive values, III. Average of force is positive and the force has both negative and positive values and IV. Average of force is positive and the force is always positive.

After proving the possibility of generating the negative averaged acoustic radiation force state as well as the positive averaged states, with or without rest states, it is time to put in challenge the suggested methodology of manipulation of activated carriers via the phase change (i.e., time varying) of incident wave field. Obviously, the dynamics and swimming quality of the carrier should be considered. Fig. 7a through Fig. 7f illustrate the time variation of velocity of carrier, $V_c$, for different values of the rate of change of the incident wave phase, defined as $V_{phase} = d\phi_0 / dt$ selected as $V_{phase}/V_{passive} = 0.01, 0.1, 1.0, 10.0, 100.0, 1000.0$ where $V_{passive}$ is the velocity of carrier in its passive state calculated as $V_{phase} = S_c E_{inc} Y_z^{passive} / (6\pi\eta a)$. For simulation, it is assumed that the carrier is in the rest state, $V_c(t=0) = 0$, the selected frequency of $ka = 2$, and selected modal phase difference, $\Delta\phi = 5.61$. As it is observed, for all rates of incident phase change, the velocity of carrier fluctuates around $S_c E_{inc} \overline{Y}_z / (6\pi\eta a)$. The range of variation, $\min(V_c) < V_c(t) < \max(V_c)$, is approximately the same for different values of $V_{phase} / V_{passive}$. The time scale of carrier velocity variation is related inversely to phase change rate, as $\propto (V_{phase} / V_{passive})^{-1}$. In conclusion, it is seen that following the suggested strategy, the carrier is manipulated with fluctuating (i.e., sinusoidal pattern) velocity around the pre-predicted velocity of swimming due to the induction of the averaged radiation force in Stokes swimming condition, keeping in mind that the low Reynolds number condition, $\rho_f a V_c / \eta \approx O(10^{-2})$, is valid. To be more specific, for a certain combination of phase difference and frequency, which those two correspond to a single value of the averaged force, the object approaches the sound source with a fluctuating velocity which the average of that velocity is independent of the rate of change of incident wave phase but increasing that rate only increases the frequency of those fluctuations.

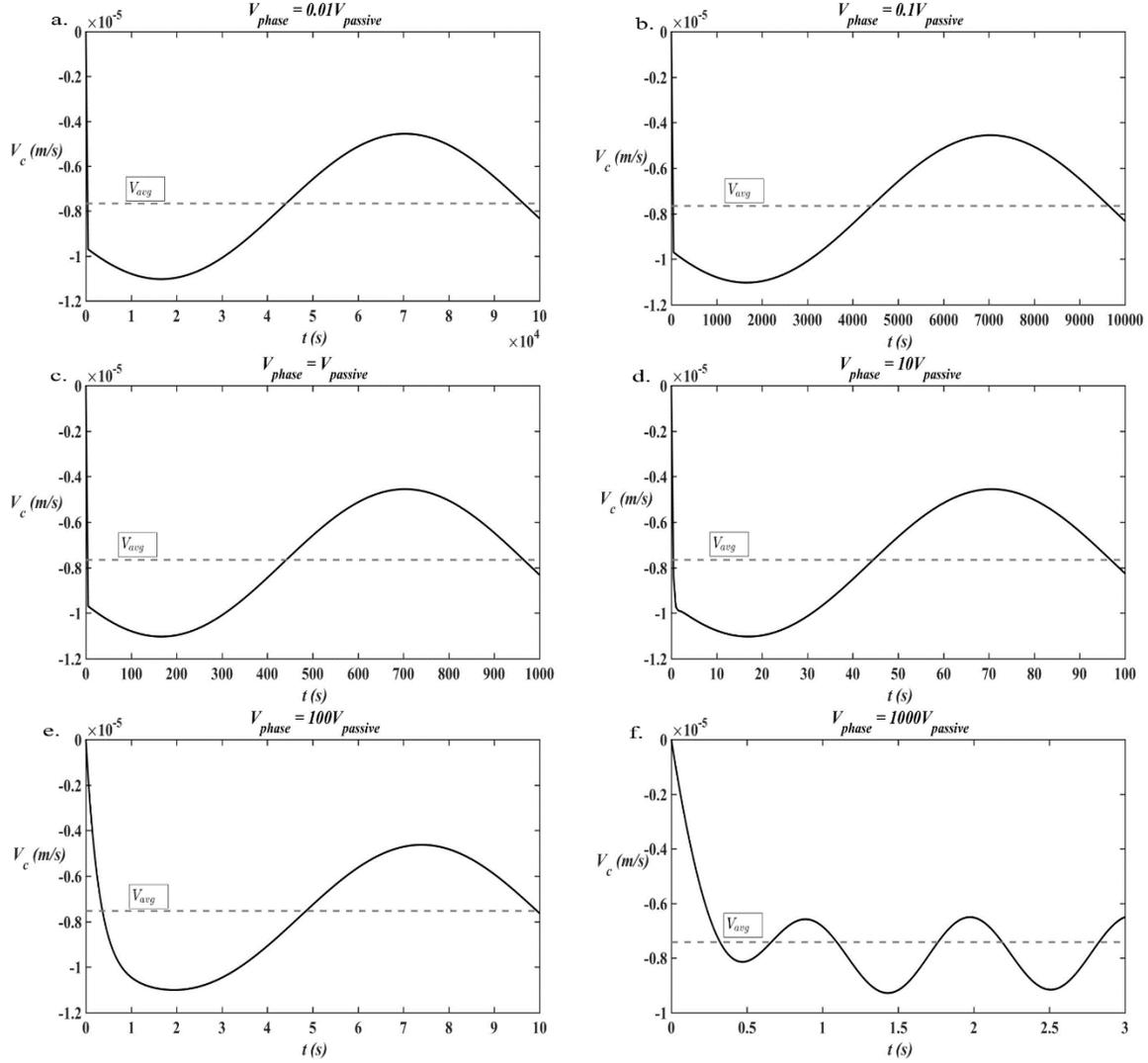

**FIG. 7.** This figure shows the velocity of particle at $ka = 2.7$ and $\Delta\varphi = 5.61\theta$ radian versus time for a sphere of radius 1 mm when the velocity of phase change is a. 0.01, b. 0.1, c. 1 d. 10, e. 100 and f. 1000 times the velocity of passive carrier.

## IV. Conclusion:

In this paper authors tried to present a simple configuration which is capable of having attractive net motion when it is illuminated by a simple acoustic progressive wave. The object here is activated by piezoelectric ceramics on its breathing and dipole modes which although the frequency of excitations is identical, they experience a difference in the phase of excitation. In the formulation it was shown that the acoustic radiation force is a function of the location of object and the average of force on a complete wavelength is a function of the phase difference between two modes. In results and discussion section it was clearly obtained that the average of force over its location is negative for a wide range of phase difference. It is notable that this work and works like this are trying to enrich the concepts and methods of complete manipulation of active carriers and practical issues should be evaluated in future works.

# Appendix

$$Z_n^1 = \frac{\begin{vmatrix} S_n^m(1,3) & S_n^m(1,4) & S_n^m(1,5) & \frac{\omega}{C_{44}}(2n+1)\rho_f\varphi_0 i^{n+1} j_n(k_f a) \\ S_n^m(2,3) & S_n^m(2,4) & S_n^m(2,5) & 0 \\ S_n^m(4,3) & S_n^m(4,4) & S_n^m(4,5) & \frac{\varphi_0}{i\omega a} k_f i^n (2n+1) j_n'(k_f a) \\ T_n^m(6,3) & T_n^m(6,4) & T_n^m(6,5) & 0 \end{vmatrix}}{\begin{vmatrix} S_n^m(1,3) & S_n^m(1,4) & S_n^m(1,5) & \frac{-\omega}{C_{44}}\rho_f\varphi_0 i^{n+1}(2n+1) h_n^{(1)}(k_f a) \\ S_n^m(2,3) & S_n^m(2,4) & S_n^m(2,5) & 0 \\ S_n^m(4,3) & S_n^m(4,4) & S_n^m(4,5) & \frac{-\varphi_0}{a\omega i} k_f i^n (2n+1) h_n^{(1)\prime}(k_f a) \\ T_n^m(6,3) & T_n^m(6,4) & T_n^m(6,5) & 0 \end{vmatrix}}, n > 0$$

$$Z_n^2 = \frac{\begin{vmatrix} S_n^m(1,3) & S_n^m(1,4) & S_n^m(1,5) & 0 \\ S_n^m(2,3) & S_n^m(2,4) & S_n^m(2,5) & 0 \\ S_n^m(4,3) & S_n^m(4,4) & S_n^m(4,5) & 0 \\ T_n^m(6,3) & T_n^m(6,4) & T_n^m(6,5) & 1 \end{vmatrix}}{\begin{vmatrix} S_n^m(1,3) & S_n^m(1,4) & S_n^m(1,5) & \frac{-\omega}{C_{44}}\rho_f\varphi_0 i^{n+1}(2n+1) h_n^{(1)}(k_f a) \\ S_n^m(2,3) & S_n^m(2,4) & S_n^m(2,5) & 0 \\ S_n^m(4,3) & S_n^m(4,4) & S_n^m(4,5) & \frac{-\varphi_0}{a\omega i} k_f i^n (2n+1) h_n^{(1)\prime}(k_f a) \\ T_n^m(6,3) & T_n^m(6,4) & T_n^m(6,5) & 0 \end{vmatrix}}, n > 0$$

where, $S_n^m(i,j)$ are the elements of the global structural transfer matrix (Rajabi and Mojahed, 2016a). It is also notable to mention that the expression for $Z_{n,m}^2$ and $Z_{n,m}^1$ for $n = 0$ is like the one for $n > 0$ with the difference that the second rows of the matrices should be eliminated.

# V. References:

A. Haake, J. D. (**2004**). "Positioning of small particles by an ultrasound field excited by surface waves," Ultrasonics **42**, 75-80.

Courtney, C. R. P., Demore, C. E. M., H. Wu, A. G., Wilcox, P. D., Cochran, S., and B. W. Drinkwater (**2014**). "Independent trapping and manipulation of microparticles using dexterous acoustic tweezers," Appl. Phys. Lett. **104**, 154103.

Foresti, D., Bjelobrk, N., Nabavi, M., and Poulikakos, D. (**2011**). "Investigation of a line-focused acoustic levitation for contactless transport of particles," Journal of Applied Physics **109**, 093503.

G. T. Silva, B. W. D. (**2018**). "Acoustic radiation force exerted on a small spheroidal rigid particle by a beam of arbitrary wavefront: Examples of traveling and standing plane waves " The Journal of the Acoustical Society of America **144**.

G. T. Silva, J. H. L., F. G. Mitri, (**2013**). "Off-axial acoustic radiation force of repulsor and tractor bessel beams on a sphere," IEEE Transactions on Ultrasonics, Ferroelectrics, and Frequency Control **60**.

Happel, J., and Brenner, H. (**1983**). *Low Reynolds Number Hydrodynamics* (Martinus Nijhoff).

Hasegawa, T. (**1977**). "Comparison of two solutions for acoustic radiation pressure on a sphere," The Journal of the Acoustical Society of America **61**, 1445-1448.

Hasegawa, T., Saka, K., Inoue, N., and Matsuzawa, K. (**1988**). "Acoustic radiation force experienced by a solid cylinder in a plane progressive sound field," The Journal of the Acoustical Society of America **83**, 1770-1775.

Hasegawa, T., and Yosioka, K. (**1969**). "Acoustic-Radiation Force on a Solid Elastic Sphere," The Journal of the Acoustical Society of America **46**, 1139-1143.

J. Happel, H. B. (**1983**). *Low Reynolds number hydrodynamics* (Springer Netherlands).

Johnson, R. W. (**1998**). *The handbook of fluid dynamics* (CRC Press).

Kozuka, T., Yasui, K., Tuziuti, T., Towata, A., and Iida, Y. (**2008**). "Acoustic Standing-Wave Field for Manipulation in Air," Japanese Journal of Applied Physics **47**, 4336.

Lim, M., Souslov, A., Vitelli, V., and Jaeger, H. (**2019a**). "Dynamics of ultrasonically levitated granular rafts," p. A59.007.

Lim, M. X., Souslov, A., Vitelli, V., and Jaeger, H. M. (**2019b**). "Cluster formation by acoustic forces and active fluctuations in levitated granular matter," Nature Physics **15**, 460-464.

Lima, E. B., Leão-Neto, J. P., Marques, A. S., Silva, G. C., Lopes, J. H., and Silva, G. T. (**2020**). "Nonlinear Interaction of Acoustic Waves with a Spheroidal Particle: Radiation Force and Torque Effects," Physical Review Applied **13**, 064048.

Liu, Y., and Hu, J. (**2009**). "Trapping of particles by the leakage of a standing wave ultrasonic field," Journal of Applied Physics **106**, 034903.

Marston, P. L. (**2006**). "Axial radiation force of a Bessel beam on a sphere and direction reversal of the force," The Journal of the Acoustical Society of America **120**, 3518-3524.

Marston, P. L. (**2007**). "Negative axial radiation forces on solid spheres and shells in a Bessel beam," The Journal of the Acoustical Society of America **122**, 3162-3165.

Marston, P. L., and Zhang, L. (**2016**). "Unphysical consequences of negative absorbed power in linear passive scattering: Implications for radiation force and torque," The Journal of the Acoustical Society of America **139**, 3139-3144.

Mitri, F. G. (**2009**). "Negative axial radiation force on a fluid and elastic spheres illuminated by a high-order Bessel beam of progressive waves," Journal of Physics A: Mathematical and Theoretical **42**, 245202.

Mojahed, A., and Rajabi, M. (**2018**). "Self-motile swimmers: Ultrasound driven spherical model," Ultrasonics **86**, 1-5.

Rajabi, M., and Hajiahmadi, A. (**2018**). "Self-propulsive swimmers: Two linked acoustic radiating spheres," Phys. Rev. E **98**, 063003.

Rajabi, M., Khodavirdi, H., and Mojahed, A. (**2020**). "Acoustic steering of active spherical carriers," Ultrasonics **105**, 106112.
Rajabi, M., and Mojahed, A. (**2016a**). "Acoustic manipulation of active spherical carriers: Generation of negative radiation force," Annals of Physics **372**, 182-200.
Rajabi, M., and Mojahed, A. (**2016b**). "Acoustic manipulation of oscillating spherical bodies: Emergence of axial negative acoustic radiation force," Journal of sound and vibration **383**, 265-276.
Rajabi, M., and Mojahed, A. (**2017a**). "Acoustic Manipulation of a Liquid-filled Spherical Shell Activated with an Internal Spherical Oscillator," Acta Acustica united with Acustica **103**, 210-218.
Rajabi, M., and Mojahed, A. (**2017b**). "Acoustic manipulation: Bessel beams and active carriers," Phys. Rev. E **96**, 043001.
Rajabi, M., and Mojahed, A. (**2018**). "Acoustic radiation force control: Pulsating spherical carriers," Ultrasonics **83**, 146-156.
S. A. Seah, B. W. D., T. Carter, R. Malkin and S. Subramanian (**2014**). "Dexterous ultrasonic levitation of millimeter-sized objects in air," IEEE Transactions on Ultrasonics, Ferroelectrics, and Frequency Control **61**, 1233-1336.
Silva, G. T. (**2011**). "An expression for the radiation force exerted by an acoustic beam with arbitrary wavefront (L) " The Journal of the Acoustical Society of America **130**, 3541-3544.
Silva, G. T. (**2014**). "Acoustic radiation force and torque on an absorbing compressible particle in an inviscid fluid," The Journal of the Acoustical Society of America **136**, 2405.
Skudrzyk, E. (**2012**). *The foundations of acoustics: basic mathematics and basic acoustics* (Springer Science & Business Media).
Torr, G. R. (**1984**). "The acoustic radiation force," American Journal of Physics **52**, 402-408.
Vessot, K. L. (**1934**). "On the acoustic radiation pressure on spheres," Proc. R. Soc. **147**, 212–240.
Westervelt, P. J. (**1951**). "The Theory of Steady Forces Caused by Sound Waves," The Journal of the Acoustical Society of America **23**, 312-315.
Yu, H.-Q., Yao, J., Wu, D.-J., Wu, X.-W., and Liu, X.-J. (**2018a**). "Negative acoustic radiation force induced on an elastic sphere by laser irradiation," Phys. Rev. E **98**.
Yu, H.-Q., Yao, J., Wu, D.-J., Wu, X.-W., and Liu, X.-J. (**2018b**). "Negative acoustic radiation force induced on an elastic sphere by laser irradiation," Physical Review E **98**, 053105.
Yu, H.-Q., Yao, J., Wu, D.-J., Wu, X.-W., and Liu, X.-J. (**2019**). "Laser irradiation modulating the acoustic radiation force acting on a liquid ball in a plane progressive wave," AIP Advances **9**, 045121.
Zhang, L., and Marston, P. L. (**2016**). "Acoustic radiation force expressed using complex phase shifts and momentum-transfer cross sections," The Journal of the Acoustical Society of America **140**, EL178-EL183.